\documentclass[10pt,Letterpaper,twocolumn]{article} 
\usepackage{ol2}
\usepackage[draft]{hyperref}
\usepackage{amsmath}
\usepackage{graphics}
\usepackage{graphicx}
\usepackage{dcolumn}
\usepackage{bm}
\usepackage{amsmath,amssymb}

\begin{document}
\twocolumn[ 
\title{Self-frequency blue-shift of dissipative solitons in silicon based waveguides}
\author{Samudra Roy$^1$, Andrea Marini$^1$ and Fabio Biancalana$^{1,2}$}
\affiliation{$^1$Max Planck Institute for the Science of Light, Guenther-Scharowsky-Stra\ss e 1, 91058 Erlangen, Germany\\
$^2$School of Engineering \& Physical Sciences, Heriot-Watt University, EH14 4AS Edinburgh, United Kingdom}
\date{\today}
\begin{abstract}
We analyze the dynamics of dissipative solitons in silicon on insulator waveguides embedded in a gain medium. The optical
propagation is modeled through a cubic Ginzburg-Landau equation for the field envelope coupled with an ordinary
differential equation accounting for the generation of free carriers owing to two-photon absorption. Our numerical
simulations clearly indicate that dissipative solitons accelerate due to the carrier-induced index change and experience
a considerable blue-shift, which is mainly hampered by the gain dispersion of the active material. Numerical results are
fully explained by analytical predictions based on soliton perturbation theory.
\end{abstract}


]

\setlength{\belowdisplayskip}{1pt} \setlength{\belowdisplayshortskip}{1pt}
\setlength{\abovedisplayskip}{1pt} \setlength{\abovedisplayshortskip}{1pt}

\noindent
Silicon photonics has attracted considerable attention among researchers owing to its potential applications ranging from
optical interconnection to bio-sensing. In the last few years, silicon-on-insulator (SOI) technology has rapidly developed
into a well-established photonic platform \cite{jalali}. The tight confinement of the optical mode and the inherently large
bulk nonlinearity of Si tremendously enhance the nonlinear dynamics \cite{leuthold}. For near-infrared wavelengths 
in the range 1 $\mu$m $< \lambda_0 < $ 2.2 $\mu$m, two photon
absorption (2PA) is the leading loss mechanism and limits the spectral broadening due to self-phase modulation (SPM) \cite{yin,husko1}.
As a consequence of 2PA, electrons are excited to the conduction band, absorb light and
affect the pulse dynamics by modifying the refractive index of silicon \cite{lin2}.
Loss mechanisms are restricted in silicon-organic hybrid slot waveguides, which can be exploited for
all-optical high-speed signal processing \cite{koos}. Alternatively, loss can be overcome by embedding active materials
in the design of SOI devices. Recently, some amplification schemes based on III-V semiconductors, rare-earth-ion-doped
dielectric thin films and erbium-doped waveguides have been proposed and practically realized \cite{fang,worhoff,agazzi}.

\begin{figure}
\begin{center}
\includegraphics[width=0.45\textwidth]{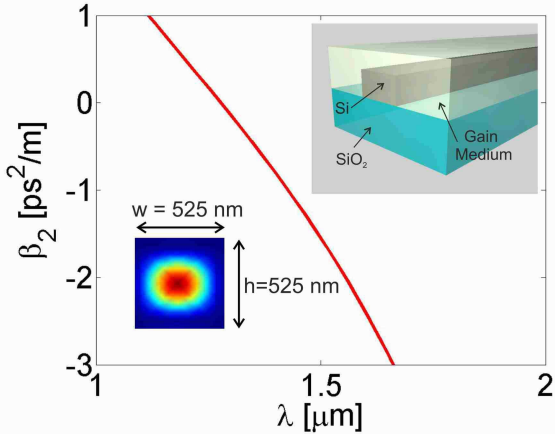}
\caption{\small{Sketch of a SOI waveguide with lateral dimensions $h = w = 525$ nm surrounded by Al$_2$O$_3$:Er$^+$
and its dispersion property. The solid red line represents the group velocity dispersion (GVD) of the quasi-TM mode,
whose spatial profile at $\lambda_0 = 1550 $ nm is depicted in the inset below.}} \label{fig1}
\end{center}
\end{figure}

In this Letter, we describe for the first time to our knowledge the nonlinear dynamics of dissipative solitons (DSs)
in amplifying SOI devices. DSs are stationary localized structures of open nonlinear systems far
from equilibrium that can be observed in several contexts \cite{akhmedievbook}. Although the model considered in our
analysis is general and can be applied to any silicon-based amplifying setup, we specialize our calculations to a
SOI waveguide embedded in Er-doped amorphous aluminium oxide (Al$_2$O$_3$:Er$^+$).  Other gain schemes involving
the use of semiconductor active materials can also be considered and gain dispersion can be reduced accordingly. For the representative structure displayed in Fig. \ref{fig1}, the gain bandwidth is of the order of $100 $ nm around the carrier wavelength $\lambda \simeq 1540 $ nm. For this waveguide,
the second order GVD coefficient and the effective area at $\lambda_0 = 1550 $ nm are calculated to be
$\beta_2 \simeq -2$ ps$^2$/m, $A_{\rm eff} \simeq 0.145$ $\mu$m$^2$, respectively. The proposed SOI waveguide is fabricated along the $[\bar{1}10]$ direction and on the $[110]\times[001]$ surface, so that quasi-TM modes do not experience
stimulated Raman scattering (SRS) \cite{lin2}. The propagation of an optical pulse with
envelope $u(z,t)$ and carrier frequency $\omega_0$ in the proposed photonic structure is governed by a complex
Ginzburg-Landau (GL) equation,
\begin{eqnarray}
&& {i} \partial_{\xi} u - \frac{1}{2}{\rm sgn}(\beta_2)\partial_{\tau}^2u + {i} \alpha u + ( 1 + {i} K ) |u|^2 u \nonumber \\
&& + ( {i}/{2} - \mu ) \phi_c u - {i} ( g + g_2 \partial_{\tau}^2 )u = 0 , \label{NLSE_DL}
\end{eqnarray}
coupled with an ordinary differential equation accounting for the 2PA-induced free-carrier dynamics
${d}\phi_c /{d} \tau = \theta |u|^4 - \tau_c \phi_c $. Eq. (\ref{NLSE_DL}) is written in dimensionless units,
where the time ($t$) and the longitudinal spatial variable ($z$) are normalized to the initial pulse width ($t=t_0\tau$) and
to the dispersion length ($z = \xi L_{\rm D}$ with $L_{\rm D}=t_0^2/|\beta_2(\omega_0)|$), respectively. The envelope amplitude
($A$) is rescaled to $A=u\sqrt{P_0}$, where $P_0=|\beta_2(\omega_0)|/(t_0^2\gamma_R)$, $\gamma_R=k_0n_2/A_{\rm eff}$ with
$n_2 \simeq ( 4 \pm 1.5 ) \times 10^{-18} $m$^2$/W is the Kerr nonlinear coefficient of bulk silicon. The bulk 2PA coefficient is
$\beta_{\rm TPA} \simeq 8 \times 10^{-12} $m/W and its corresponding effective waveguide counterpart
$\gamma_I=\beta_{\rm TPA}/(2A_{\rm eff})$ is rescaled to the Kerr coefficient so that
$K=\gamma_I/\gamma_R=\beta_{\rm TPA}\lambda_0/(4\pi n_2)$. The linear loss coefficient ($\alpha_l$) is renormalized to the
dispersion length ($\alpha = \alpha_l L_{\rm D}$) and can be neglected for short propagation in the linear transparency spectral window of silicon
1 $\mu$m$ < \lambda_0 < $10 $\mu$m, where 2PA dominates (if $\lambda_0 < 2.2$  $\mu$m). The density of free-carriers (FCs)
$N_c$ generated through 2PA is normalized so that $\phi_c = \sigma N_c L_{\rm D}$, where
$\sigma \simeq 1.45 \times 10^{-21} $m$^2$. FCs are responsible for free-carrier dispersion (FCD)
regulated by the parameter $\theta=\beta_{\rm TPA}|\beta_{2}|\sigma/(2 \hbar\omega_0A_{\rm eff}^2t_{0}\gamma_{R}^2)$ \cite{lin}
and free-carrier absorption (FCA) depending on the parameter $\mu=2\pi k_{c}/(\sigma\lambda_{0})$, where
$k_c\simeq1.35\times10^{-27} $m$^3$ \cite{Dinu}. $t_c$ is the characteristic FC recombination time (of the order of ns) and
normalized as $\tau_c=t_0/t_{c}$, which we neglect in our calculations since we focus our analysis
on ultrashort pulses with time duration  of the order $t_0 \simeq 100$ fs. The amplifying medium is characterized by a
gain coefficient $G$, which in our dimensionless equations is rescaled to the dispersion length ($g = G L_{\rm D}$),
and a dephasing time $T_2$, which is related to the dimensionless gain dispersion coefficient through $g_2 = g (T_2/t_0)^2$.

The presence of gain and 2PA in the photonic structure considered in our calculations implies that, in  general, energy is not
conserved. The GL equation does not support solitons in a strict mathematical sense since it is not integrable by the
inverse scattering transform. In contrast to Kerr solitons in conservative systems, which form continuous families of localized
solutions, DSs are formed under a dynamical equilibrium involving non-trivial internal energy flows. DSs are associated with certain discrete parameters of the GL equation that satisfy the energy balance condition. In absence of linear loss (i.e. $\alpha=0$) and
free-carriers (i.e. $\phi_c=0$), DSs can be found as
$u(\xi,\tau)=u_{0}[{\mathrm sech}(\eta\tau)]^{1-{i}\beta}{e}^{{i}\Gamma\xi}$,
where the exact values of $u_0,\eta,\beta,\Gamma$ are fixed by the physical parameters of the system $g,g_2$ and $K$ \cite{agrawalbook}.
This soliton suffers from inherent core and background instabilities and can be stabilized by introducing higher order
nonlinearities or by coupling the system to a passive waveguide \cite{malomed}. However, this task goes beyond the scope of the
present work, where we aim at understanding the effect of FCs on the DS dynamics. In order to grasp the fundamental effects induced
by FC dynamics, we develop a soliton perturbative analysis \cite{hasegawabook}, approximating the GL equation as a perturbed
nonlinear Schr\"odinger equation (NLSE): ${\mathrm i}\partial_\xi u+\frac{1}{2}\partial_\tau^2 u+|u|^2 u={i}\epsilon(u)$,
where we explicitly consider the case of anomalous dispersion ($\beta_2<0$) and $\epsilon(u)$ includes the coupling to FCs, 2PA,
linear gain and its dispersion: $\epsilon=(g-K|u|^2-\phi_c/2-{i}\mu\phi_c+g_2\partial_{\tau}^2)u$. The perturbative theory
is developed by making the {\it Ansatz}:
\begin{eqnarray}
u(\xi,\tau) & = & u_0(\xi) \left[ {\mathrm sech} \left\{\eta(\xi)[\tau-\tau_{p}(\xi)]\right\} \right]^{1-{ i}\beta} \times \nonumber \\
            &   &  {e}^{{i}\phi(\xi)-{i}\Omega(\xi)(\tau-\tau_p(\xi))}, \label{ansatz}
\end{eqnarray}
where the parameters $u_0,\eta,\tau_p,\phi,\Omega$ are now assumed to depend on $\xi$ and $\epsilon(u)$ is considered as a small
perturbation depending on $u$, $u^*$ and their derivatives. The evolution dynamics of the soliton parameters over distance
can then be predicted using the variational method \cite{agrawalbook}, which leads to a set of coupled differential equations for
the soliton parameters. The evolution of pulse energy ($E$), frequency ($\Omega$) and temporal ($\tau_p$) shifts
are given by
\begin{eqnarray}
&& \frac{{d} E}{{d} \xi} = - 2 E \left(\frac{1}{6} \theta u_0^2 E  + g_2 \Omega^2 \right) , \label{E} \\
&& \frac{{d} \Omega}{{d} \xi} = \frac{8}{15} (\mu+\beta/2) \theta u_0^4 - \frac{4}{3} g_2 (1+\beta^2) \Omega \eta^2 , \label{Frq} \\
&& \frac{{d} \tau_p}{{d} \xi}= - (1 - 2 g_2 \beta ) \Omega - \frac{7}{72}\theta E^2 . \label{position}
\end{eqnarray}

\begin{figure}
\centering
\begin{center}
\includegraphics[width=0.235\textwidth]{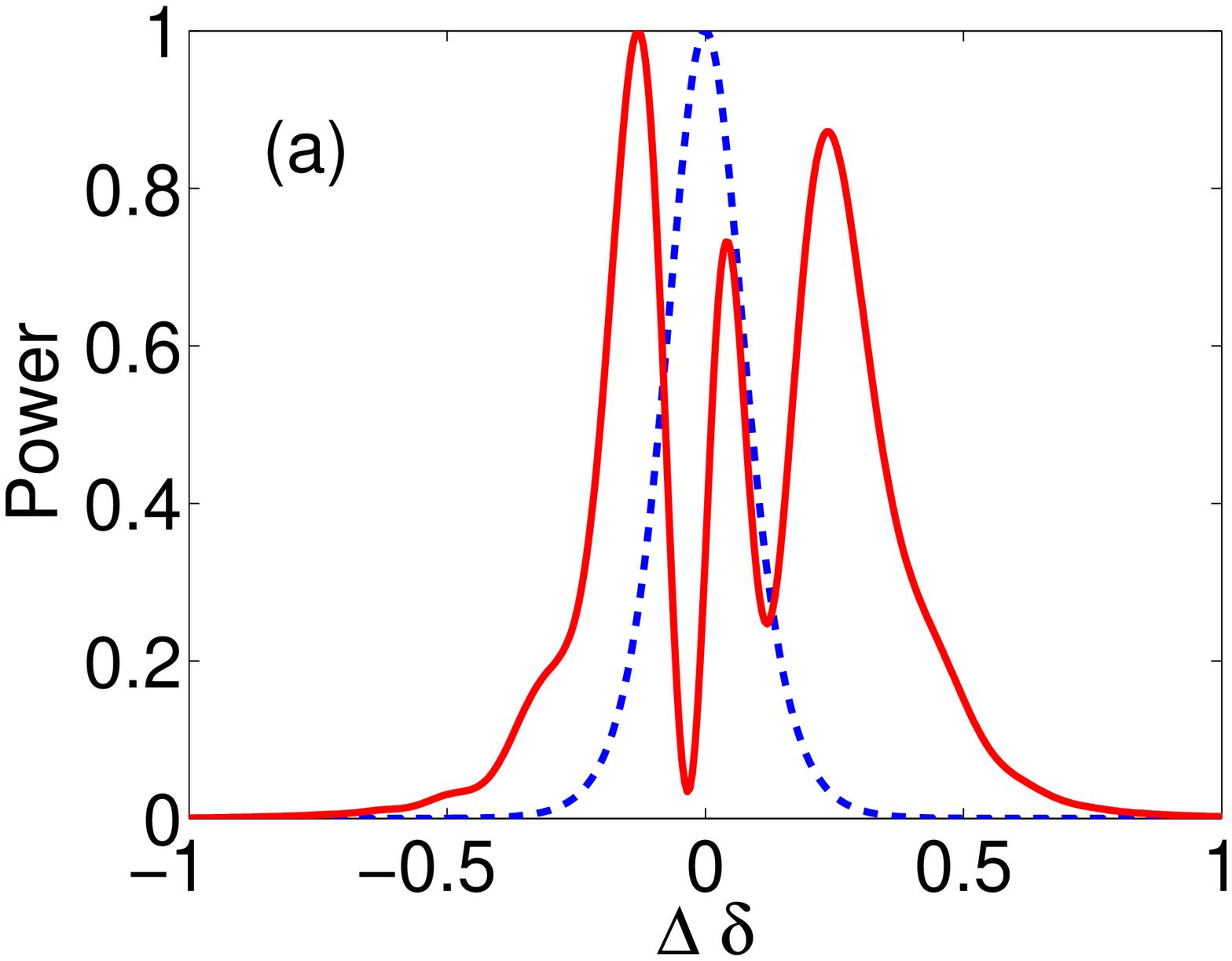}
\includegraphics[width=0.235\textwidth]{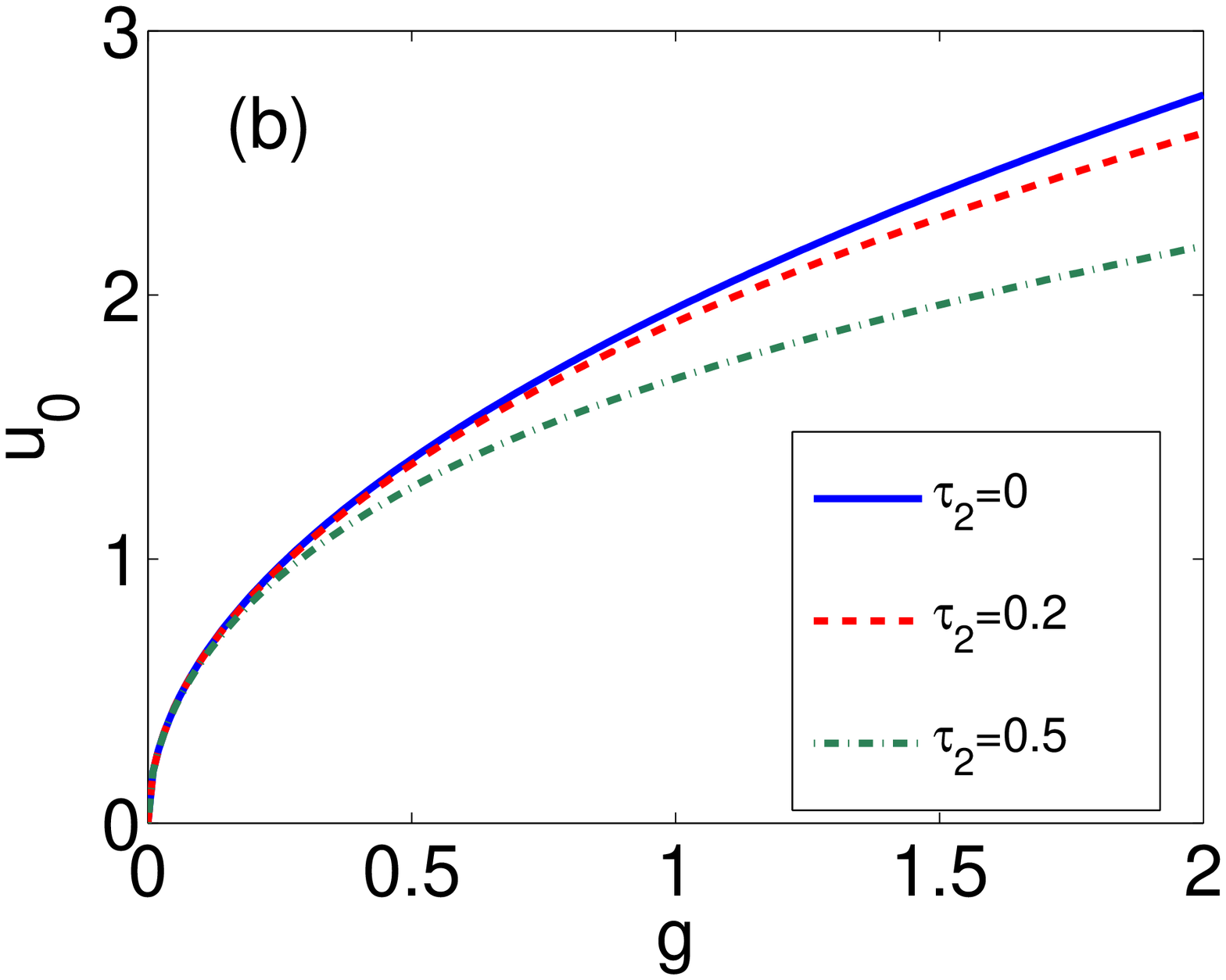}
\includegraphics[width=0.235\textwidth]{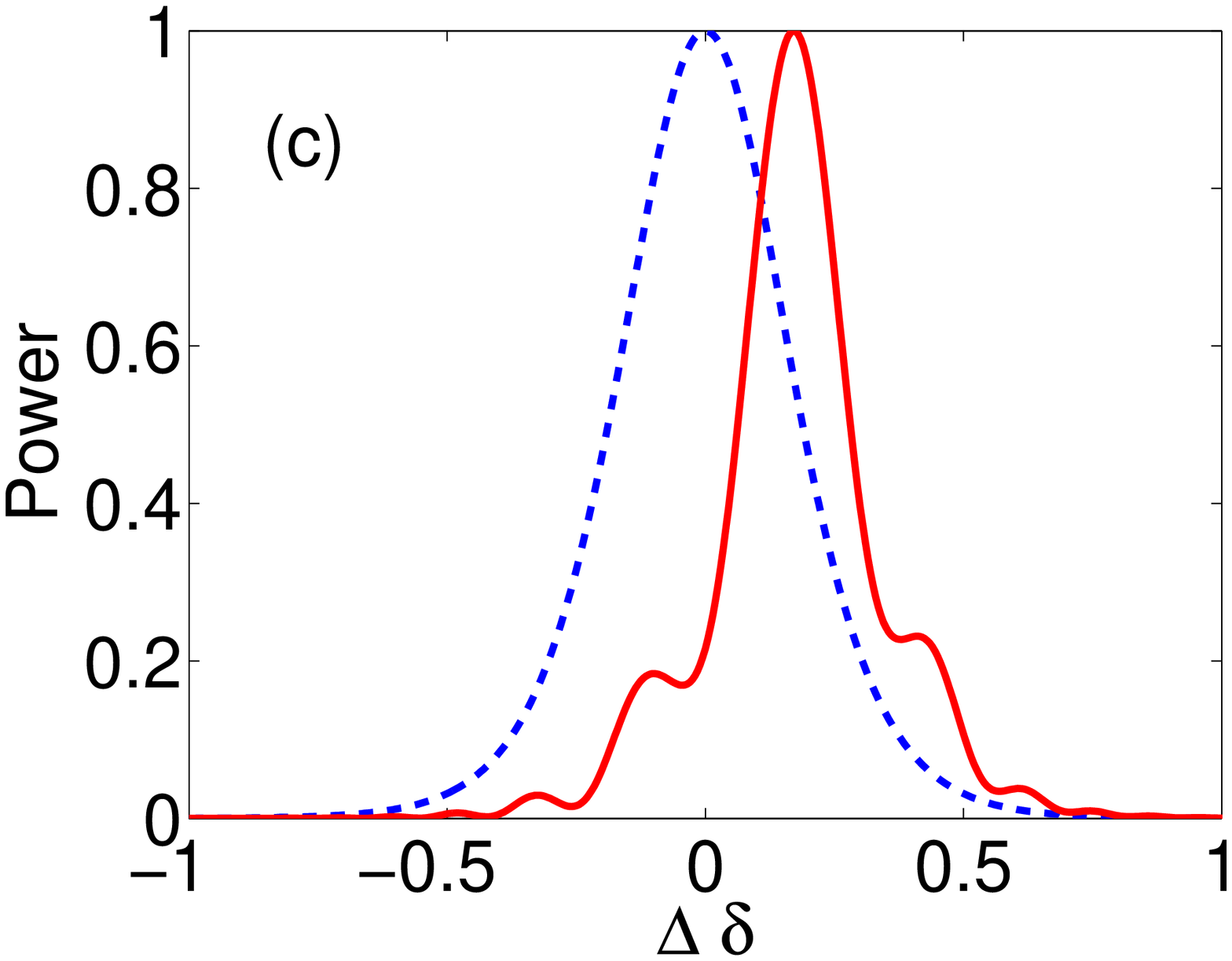}
\includegraphics[width=0.235\textwidth]{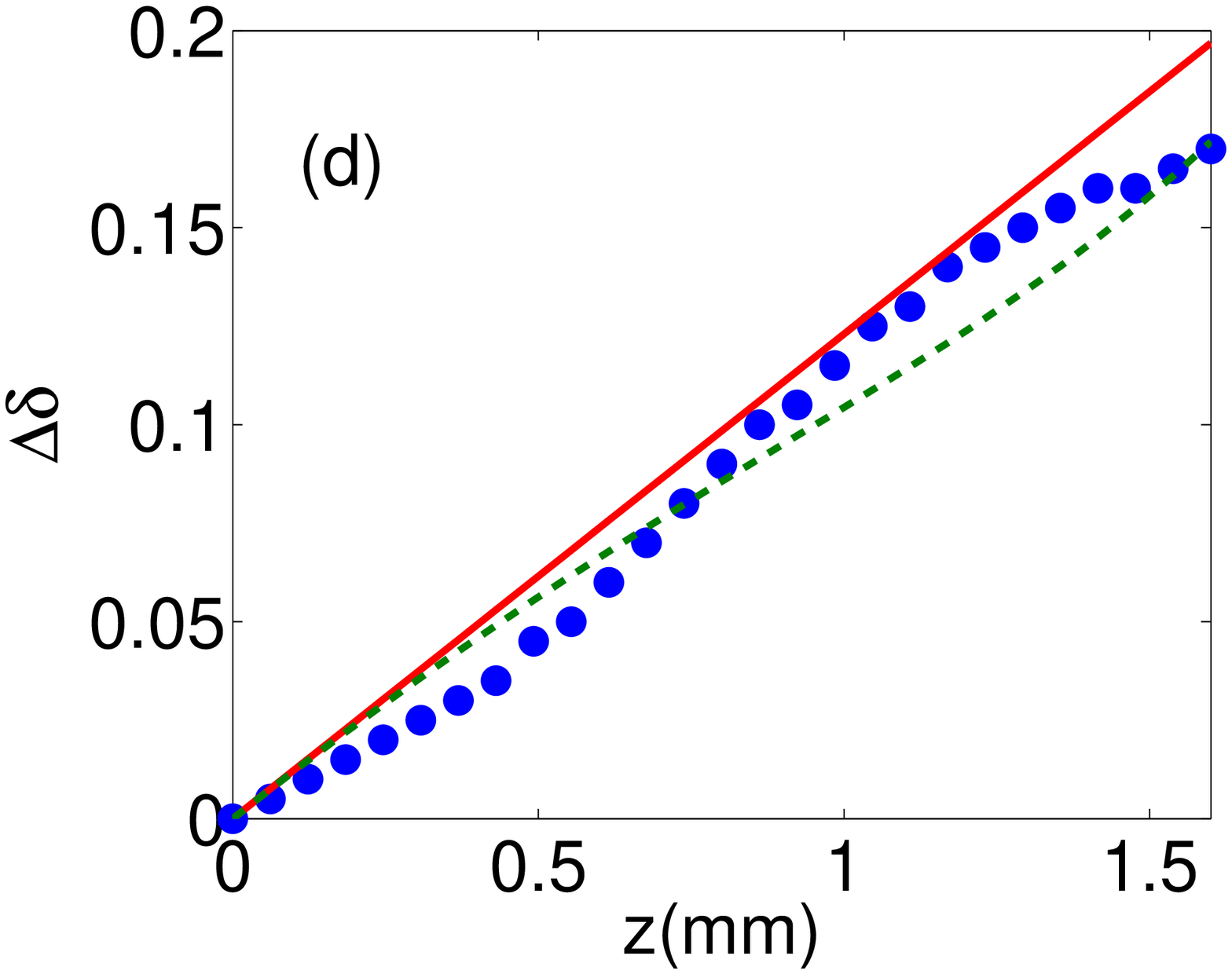}
\caption{\small{(a) Asymmetric spectral broadening of the NLSE soliton ($u_{in} = {\mathrm sech} \tau$) with time duration $t_0 = 40$ fs for
$g=1$ and $g_2=0$. (a) DS amplitude $u_0$ as a function of linear gain $g$ for several dephasing times ($\tau_2=T_2/t_0 = 0,0.2,0.5$) at fixed $K$.
(c) Carrier-induced blue-shift of DS ($u_{in}=u_0[{\mathrm sech}(\eta\tau)]^{1-{\mathrm i}\beta}$) for the same parameters of (a), where $u_0=\sqrt{3g/(2K)}$, $\eta=\sqrt{-g/\beta}$ and $\beta=3/(2K)-\sqrt{[3/(2K)]^2+2}$.
Solid red and blue dashed lines indicate the output and input power spectra, respectively. (d) $z$-dependent DS
frequency shift ($\Delta \delta = \Omega/(2\pi)$) for $g_2=0$. Perturbative predictions with constant (solid red line)
and $z$-dependent (green dotted line) peak amplitude $u_0$ are shown. The solid red dots indicate numerical findings.}}
\end{center}
\end{figure}

\noindent The perturbative analysis reveals that FCs induce frequency blue-shift and temporal acceleration of DSs, both effects
being hampered by the gain dispersion that limits blue-shifting within the amplifying frequency window of the active material.
For small gain dispersion, Eq. (\ref {Frq}) is solved with the initial condition $\Omega(0)=0$, obtaining $\Omega(\xi)= f\xi$,
where $f=8(\mu+\beta/2)\theta u_0^4/15$. This expression predicts that DSs experience a spectral blue-shift proportional to the
propagation distance through a rate ($f$) that basically depends on the FC density. The equation of the temporal shift turns out to be $\tau_p(\xi)=-(a\xi+f\xi^2/2)$, where $a=7\theta E^2/72$. The expression suggests during the propagation DS is accelerated under the influence of FCs analogously to the recently studied case of gas-filled hollow-core photonic crystal fibers\cite{saleh}.

\begin{figure}
\centering
\begin{center}
\includegraphics[width=0.235\textwidth]{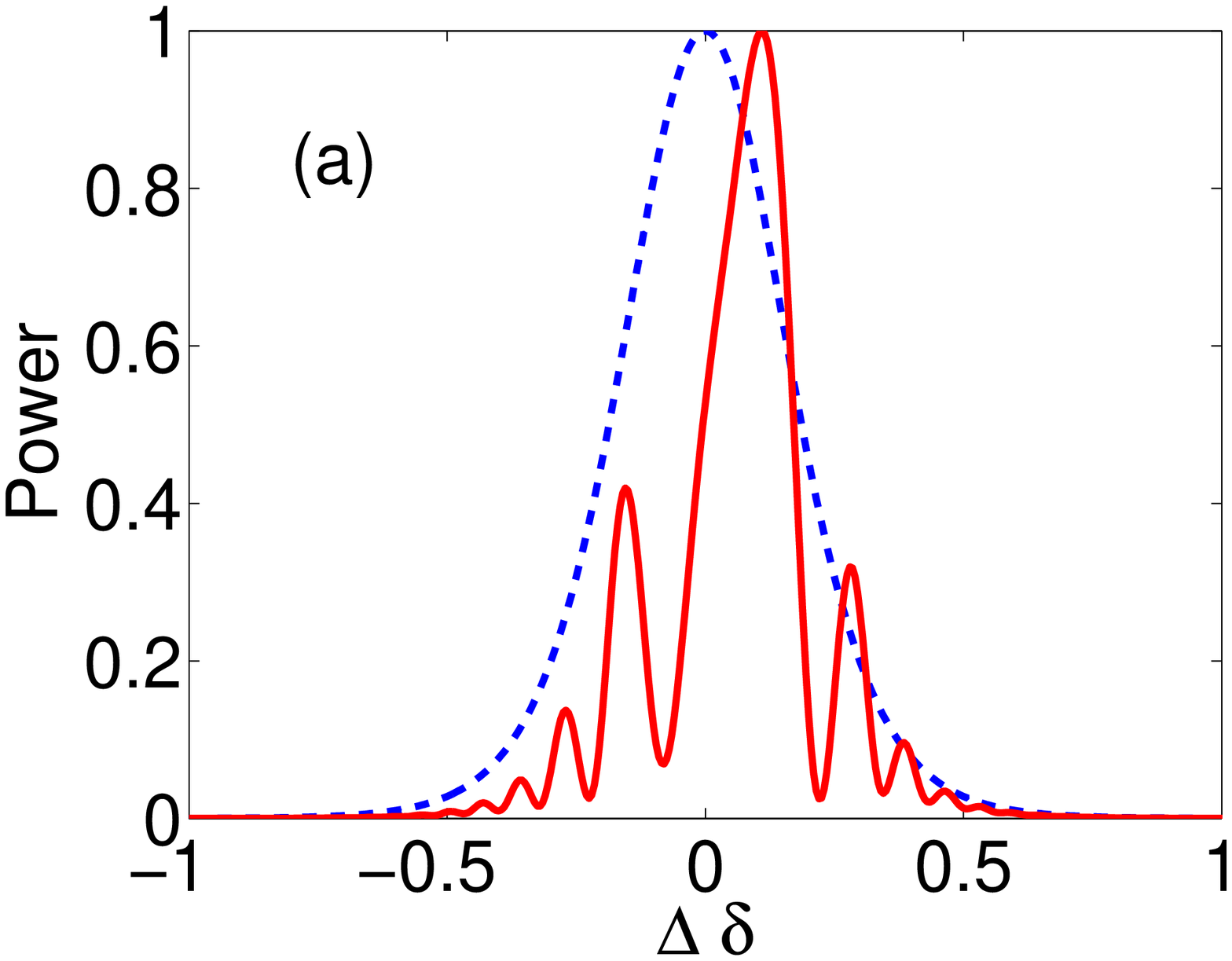}
\includegraphics[width=0.235\textwidth]{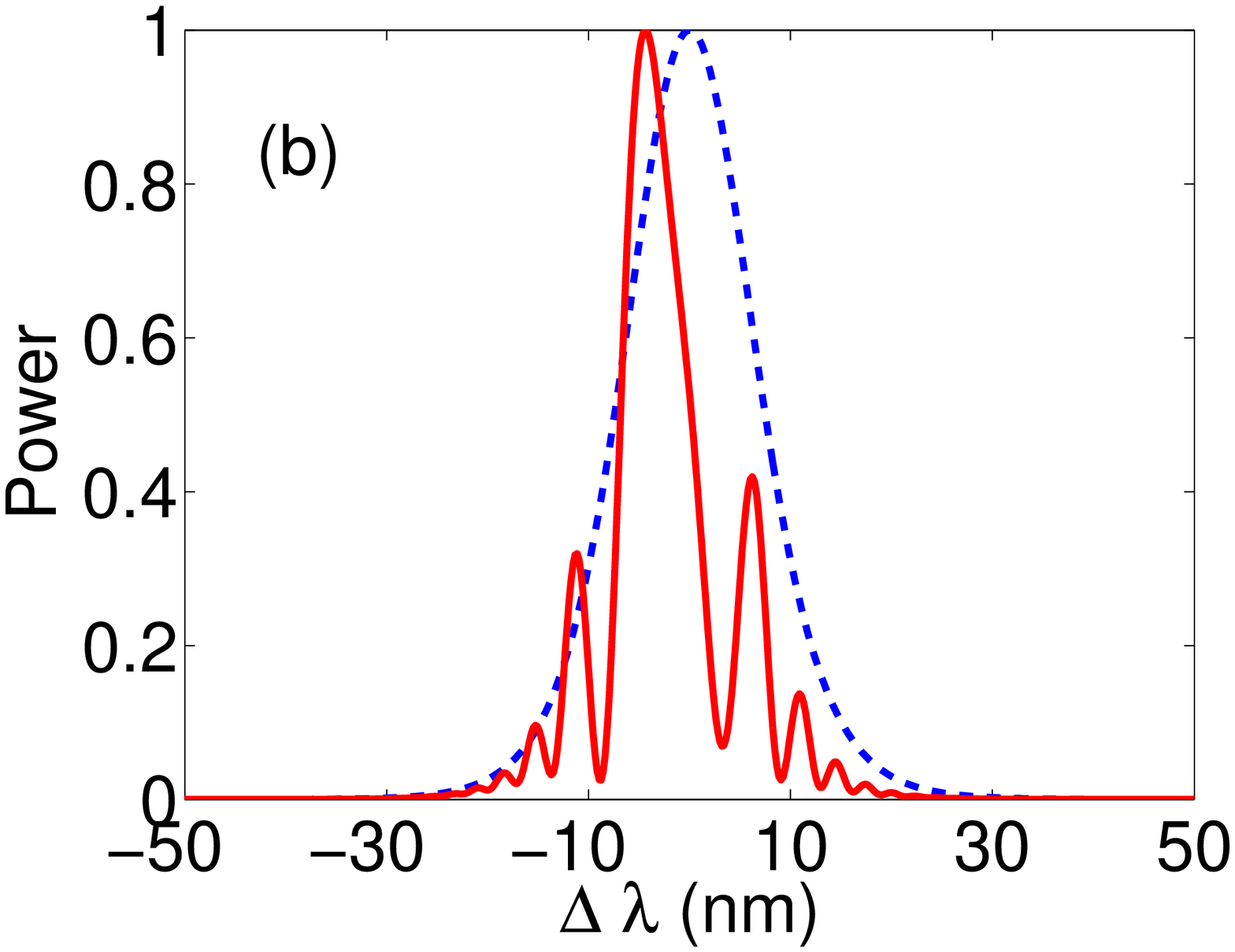}
\includegraphics[width=0.235\textwidth]{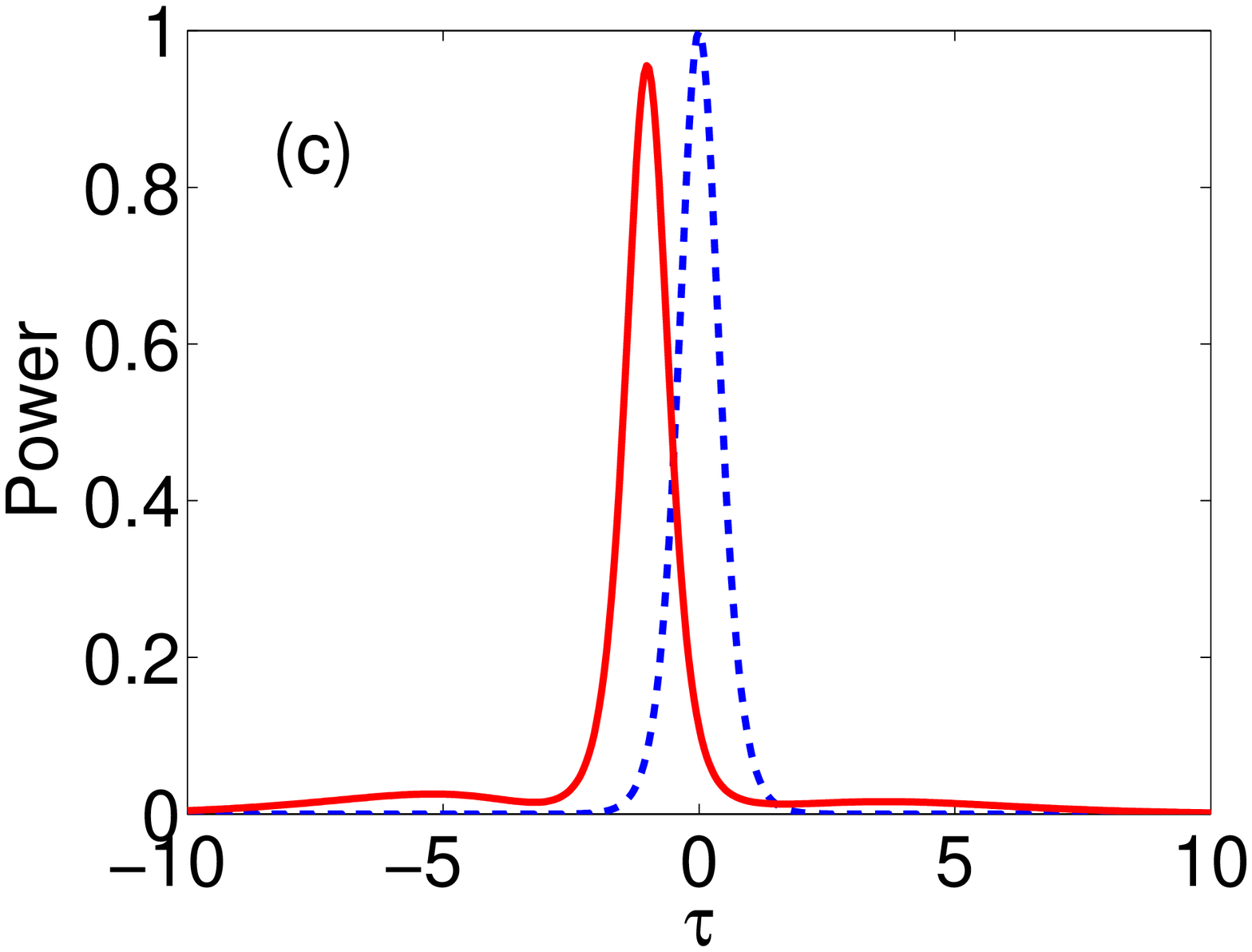}
\includegraphics[width=0.235\textwidth]{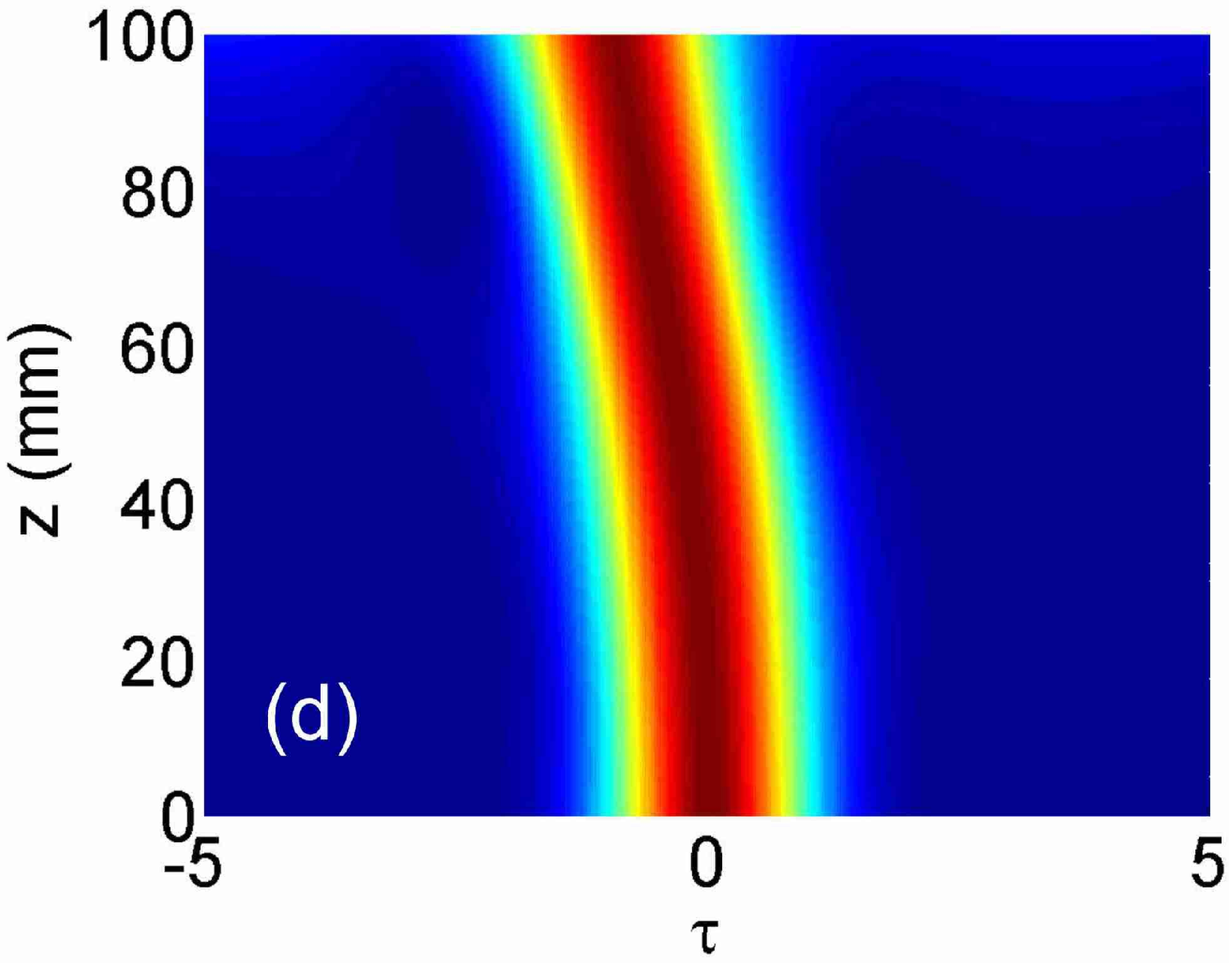}
\caption{\small{(a,b) Input (dashed blue line) and output ($\xi=5$, full red line) power spectrum of a dissipative soliton
with carrier wavelength $\lambda_0 = 1550 $ nm and time duration $t_0=200 $ fs for $g=1,g_2=0.04$ as a function
of (a) dimensionless frequency and (b) physical wavelength. (c) Input (dashed blue line) and output ($\xi=5$, full red line)
temporal DS profile. (d) Intensity counterplot of the spatio-temporal evolution of an accelerated dissipative soliton.}}
\end{center}
\end{figure}

In Fig 2a we show the conventional asymmetric spectral broadening of NLSE solitons \cite{husko2}. In Fig. 2b we plot the DS amplitude ($u_0$) as a function of linear gain for several dephasing times ($\tau_2=T_2/t_0 = 0,0.2,0.5$).
As shown in Fig. 2c, even though DSs of the cubic GL equation are unstable, they experience a considerable carrier-induced blue-shift
maintaining their shape over a propagation distance of the order of millimeters.  Indeed, for DSs, 2PA is exactly compensated by the linear gain, while NLSE solitons experience an imbalanced dynamical evolution due to 2PA and amplification.
 In Fig. 2d we compare the perturbatively predicted FC-induced frequency blue-shift (full blue and dashed green lines) with numerical results (solid red dots), finding that for the initial part of pulse propagation, perturbative predictions nicely match with the numerical findings. The dimensionless frequency shift is calculated to be $\Delta\delta \simeq 0.16$ at $\xi=2$, corresponding in physical units to $\Delta \lambda \simeq35 $ nm. The solid red line in Fig. 2d
represents the predicted frequency shift calculated through the perturbation analysis by approximating the DS amplitude to remain
constant, whereas the green dotted line represents the perturbative prediction considering $z$-dependent DS amplitudes.
From the very beginning of the propagation, the pulse shape is perturbed by FCD and FCA and hence the approximate analytic treatment fails for long propagation distances.
Fig. 3 displays the effect of gain dispersion over the pulse propagation. Due to the finite bandwidth of the amplifying medium, the frequency
shift is hampered after some saturation frequency ($\Omega_{sat}$), as clearly predicted by our perturbative analysis:
$\Omega(\xi) \simeq \Omega_{sat} (1-{\mathrm e}^{-\rho\xi})$, where $\Omega_{sat} = ( \mu + \beta / 2 ) \theta E^2 / [10 g_2(1+\beta^2)]$
and $\rho = ( 4 / 3 ) g_2 ( 1 + \beta^2 ) \eta^2$. In Figs. 3a,b we depict the input and output spectral power of a DS in presence
of gain dispersion indicating that blue-shift is reduced, as predicted by the theory. In Fig. 3c we show the temporal profile of a
propagating pulse at $\xi=5$ where the inherent background instability of DSs starts to affect the pulse propagation. The spatio-temporal
pulse evolution is displayed in Fig. 3d, where the pulse acceleration is reduced by the effect of gain dispersion.

In conclusion, we have provided a complete theoretical analysis of the the carrier-induced DS dynamics in Si-based waveguides
embedded in an amplifying medium. FCs generated through 2PA affect the refractive index of the medium and lead to a considerable
self-frequency blue-shift. We have derived analytical predictions for the self-frequency blue-shift and temporal evolution based
on soliton perturbative theory. We also examined the fully realistic condition where gain dispersion hampers the continuous
spectral blue-shifting, which is still observed.

\indent
\indent

This research was funded by the German Max Planck Society for the Advancement of Science (MPG).

\end{document}